\documentclass[12pt]{article}
\usepackage{natbib}
\usepackage{authblk}
\usepackage{amsmath,amsfonts}
\usepackage{graphicx}
\usepackage{bbm}
\usepackage{adjustbox}

\newcommand{\Pro}{\mathbb{P}}
\newcommand{\X}{\boldsymbol{X}}
\newcommand{\Y}{\boldsymbol{Y}}
\newcommand{\I}{\boldsymbol{I}}

\newcommand{\ind}{\mathbbm{1}}
\newcommand{\N}{\mathbb{N}}
\newcommand{\R}{\mathbb{R}}

\newcommand{\indep}{\perp \!\!\! \perp}

\title{A regularized MANOVA test for semicontinuous high-dimensional data}

\author[1]{Elena Sabbioni}
\author[2]{Claudio Agostinelli}
\author[3]{Alessio Farcomeni}

\affil[1]{Politecnico di Torino, Dpt. of Mathematical Science, \texttt{elena.sabbioni@polito.it}}
\affil[2]{University of Trento, Dpt. of Mathematics, \texttt{claudio.agostinelli@unitn.it}}
\affil[3]{Tor Vergata University of Rome, Dpt. of Economics and Finance, \texttt{alessio.farcomeni@uniroma2.it}}

\begin{document}

\maketitle
\begin{abstract}
We propose a MANOVA test for semicontinuous data that is applicable also when the dimensionality exceeds the sample size. The test statistic is obtained as a likelihood ratio, where numerator and denominator
are computed at the maxima of penalized likelihood functions under each hypothesis. Closed form solutions for the regularized estimators allow us to avoid computational overheads. We derive the null distribution using a permutation scheme. The power and level of the resulting test are evaluated in a simulation study.
We illustrate the new methodology with two original data analyses, one regarding microRNA expression in human blastocyst cultures, and another regarding alien plant species invasion in the island of Socotra (Yemen). 
\noindent \textbf{Keywords}: Penalized methods, Permutation tests, Zero-inflation.
\end{abstract}

\section{Introduction}
\label{sec1}

A well known generalization of ANOVA tests for the case of multivariate data is the MANOVA test\citep{wilks}.
The test involves a joint null hypothesis claiming homogeneity of mean vectors of two or more multivariate Gaussian distributions,
usually under an assumption of homoskedasticity. While classical tests are devised for low dimensional settings, 
there are several contributions for continuous data when the dimensionality exceeds the sample size, with recent reviews \citep{harr:kong:22}.
More in detail, some works involve some form of regularization \citep{chen2011regularized, cai:xia:14, dong:et:al:17}, in order to guarantee positive definiteness of the (possibly common) variance covariance matrix;
other works involve estimation of the traces of covariance matrices \citep{chen:qin:10,sriv:kubo:13,yama:hime:15, hu:et:al:17}.

In this work we are interested in semicontinuous data, that involve positive continuous measurements plus a certain fraction of exactly zero observations. Semicontinuous data are measured in several important applications like abundance estimation in ecology \citep{farcomeni2016manova}, omics data \citep{taylor2009hypothesis}, cost-effective analysis in medical research \citep{tu1999wald}.  
Inference involves separately modeling the mass probability at zero, and the positive continuous observations conditionally on a non-zero measurement \citep{lach:01, lach:02, chai:bail:08}. A common assumption is that the positive measurements are log-normal. 
Tests have appeared for different formulations of the null hypothesis \citep{xiao:tu:99, tu1999wald, xiao1999comparison},
which inevitably involve both the occurrence probabilities and the conditional moments of the continuous part. 
Formal MANOVA tests for semicontinuous data have appeared for
the case in which the dimensionality does not exceed the sample size \citep{farcomeni2016manova}.
Here we propose a MANOVA test for semicontinuous data, based on regularization,
that is applicable also when the dimensionality exceeds the sample size.
This is a quite common occurrence in modern applications, including the areas discussed above. To motivate our contribution for the applied readership,
we present two original data examples. One involves testing homogeneity of microRNA expression in human blastocyists, where many zeros are observed
due to the very early stage of development, and the dimensionality is large. The other application involves comparing overall abundance of
several alien species in the island of Socotra (Yemen), where many alien species might not be observed in a few sampling spots.
To the best of our knowledge, our MANOVA test for high dimensional semicontinous data is the first test of this kind to appear in the literature.

The rest of the paper is as follows: in the next section we set up the general framework and notation. A penalized version of the
maximum likelihood estimator for the general model is described in Section \ref{subsec:likelihood inference}. The null hypothesis
and the resulting likelihood-ratio type test are describe in Section \ref{sec:test}. The resulting test is evaluated
through a simulation study in Section \ref{sec:sim}, where we report on the observed level and power over different scenarios.
We then illustrate the new methodology through two original real data examples. The microRNA data are described and analyzed in Section \ref{subsec:microRNA};
while the alien species data in Section \ref{subsec:alien}. Some concluding remarks are given in Section \ref{sec:concl}. 

The methods discussed in this paper have been implemented in an {\tt R} package called {\tt semicontMANOVA},
whose source and compiled version are available along with this paper on the publisher's website.
\section{General framework}

Assume we have $K$ groups, each with $n_k$ observations, where $k=1, \dots, K$. Let $n := \sum_{k=1}^K n_k$ denote the total number of observations. Let $\X_k$ be a $n_k \times p$ matrix that contains $n_k$ $p$-dimensional observations of the $k$-th group.
In our framework each element $X_{ijk}$ can be either zero, or positive, and we expect a positive proportion of zeros.

Consider $Y_{ijk}:= \ind(X_{ijk}>0)$, where $\ind(\cdot)$ is the indicator function and let $\boldsymbol{a}:=(a_1, \dots, a_p) \in \{0,1\}^p$ be any of the possible configurations of presences ($Y_{ijk} = 1)$ and absences ($Y_{ijk} = 0$) over $p$ components. We model $\Y_{ik}=\left(Y_{i1k}, \dots, Y_{ipk}\right)$ with a multivariate Bernoulli distribution, that is fully specified by $2^{p}-1$ free parameters of the kind $\pi_k(\boldsymbol{a}):= \Pro\left(Y_{i1k}=a_1, \dots, Y_{ipk}=a_p\right)$, assuming $\sum_{\boldsymbol{a}\in \{0,1\}^p} \pi_k(\boldsymbol{a}) = 1$. 

For all $k=1, \ldots, K$, $i=1, \ldots, n_k$ and $j=1, \ldots, p$, we let $\Tilde{X}_{ijk} := \log{X_{ijk}}$ when $Y_{ijk} = 1$, otherwise $\Tilde{X}_{ijk}$ is not observed. We assume $\Tilde{\X}_{ik}$ follows a multivariate $p$-dimensional Gaussian distribution, with mean $\boldsymbol{\mu}_k$ and covariance matrix $\boldsymbol{\Sigma}$ when all the components are observed. We thus make an homoskedasticity assumption.

Estimating all the parameters of the Bernoulli distribution is reasonable when $p$ is small, while their number is definitely not manageable in situations of high dimensionality of the data, especially when $p > n$.
We reduce the number of parameters by assuming that the joint probability $\pi_k$ depends only on the number of positive components that are present in one observation. Formally, we assume 
\begin{equation}
	\label{eq:assumption_pi}
	\pi_k(\boldsymbol{a}_r) = \pi_k(\boldsymbol{a}_t) \quad \text{if } \boldsymbol{a}_r, \boldsymbol{a}_t \in \{0,1\}^p \text{ s.t. } \sum_{j=1}^p a_{jr} = \sum_{j=1}^p a_{jt}.
\end{equation}
To simplify the notation, we define $\pi_k(s):= \pi_k(\boldsymbol{a})$ for each $\boldsymbol{a}$ such that $s:= \sum_{j = 1}^p a_j$, where $s = 0, \dots, p$. Assuming \eqref{eq:assumption_pi}, the number of parameters describing the presence/absence part of the data decreases from $K(2^p-1)$ to $Kp$. Assumption \eqref{eq:assumption_pi} above can be verified up front through a simple chi-square test. We let $\boldsymbol{\theta} := \left( \boldsymbol{\pi}, \boldsymbol{\mu}, \boldsymbol{\Sigma} \right)$ be the set of all unknown parameters in the parametric space $\Theta = (0,1)^{Kp} \times \mathbb{R}^{Kp} \times S_p$, where $S_p$ is the set of all symmetric positive definite $p\times p$ matrices. The total number of parameters is $\vartheta = 2Kp+p(p+1)/2$.

\subsection{Likelihood inference}
\label{subsec:likelihood inference}

In this section we introduce a penalized version of the Maximum Likelihood Estimator (MLE) for all the parameters $\boldsymbol{\theta}$.
The penalty is necessary for the case in which $p>n$, and might be useful otherwise (e.g., an increased bias might correspond to a decrease in variance). We define the set $V(\Y_{ik}):= \{j \in \{1, \dots, p\}: Y_{ijk}=1 \}$, containing the indices of all positive (observed) components of $\X_{ik}$. In the following discussion, given a $p$-dimensional vector $\boldsymbol{b}$ and a set $V \subseteq \{1, \dots, p\}$, $\boldsymbol{b}_V$ is the $|V|$-dimensional vector with the components of $\boldsymbol{b}$ that are indexed by the elements of $V$. Similarly, given a $p \times p$ matrix $A$, $A_V$ indicates the $|V| \times |V|$ matrix built considering only the rows and the columns of $A$ that have indices belonging to $V$. Hence, let $n_{ik}:= \sum_{j=1}^p Y_{ijk}$ the number of observed components for the $i$th observation in the $k$th group. We have
\begin{equation}
	\label{Xtilde}
	\Tilde{\X}_{ik}|\Y_{ik} \sim \mathcal{N}_{n_{ik}} \left(\boldsymbol{\mu}_{V(\Y_{ik})k}, \boldsymbol{\Sigma}_{V(\Y_{ik})} \right),
\end{equation}
i.e., conditionally on the presence-absence vector $\Y_{ik}$, the logarithm of the positive components of $\X_{ik}$ follows a multivariate Gaussian distribution, with dimension equal to the number of non-zero components and parameters obtained as the opportune subset of the $p$-dimensional parameters $\boldsymbol{\mu}_k$ and $\boldsymbol{\Sigma}$. Combining the information from $\Tilde{\X}_{ik}$ and $\Y_{ik}$, the likelihood is given by 
\begin{align*}
	\mathcal{L}(\boldsymbol{\theta}) & = \prod_{k=1}^K \prod_{i=1}^{n_k} \prod_{\boldsymbol{a}} \pi_k(\boldsymbol{a})^{\ind(V(\boldsymbol{a})=V(\mathbf{Y}_{ik}))} \times \prod_{k=1}^K \prod_{i=1}^{n_k} \left\{ (2\pi)^{-\frac{n_{ik}}{2}}  \left|\boldsymbol{\Sigma}_{V(\mathbf{Y}_{ik})} \right|^{-\frac{1}{2}} \right. \\
	& \times \left. \exp\left[ -\frac{1}{2} \left(\Tilde{\X}_{iV(\mathbf{Y}_{ik}) k } - \boldsymbol{\mu}_{V(\mathbf{Y}_{ik})k} \right)^\top \boldsymbol{\Sigma}_{V(\mathbf{Y}_{ik})}^{-1} \left(\Tilde{\X}_{i V(\mathbf{Y}_{ik}) k } - \boldsymbol{\mu}_{V(\mathbf{Y}_{ik})k}\right) \right] \right\} \ .
\end{align*}
and the correspoding log--likelihood is
\begin{equation*}
	\begin{split}
		\ell(\boldsymbol{\pi}, \boldsymbol{\mu}, \boldsymbol{\Sigma})
		& = \sum_{k=1}^K \sum_{i=1}^{n_k} \sum_{\boldsymbol{a}} \ind(V(\boldsymbol{a})=V(\mathbf{Y}_{ik}))
		\log(\pi_k(\boldsymbol{a})) + \sum_{k=1}^K \sum_{i=1}^{n_k} \Bigg[ \frac{-n_{ik}}{2}\log(2\pi)+\\
		&-\frac{1}{2}\log\left(\left|\boldsymbol{\Sigma}_{V(\mathbf{Y}_{ik})} \right|\right)+ \\
		&-\frac{1}{2} \left(\Tilde{\X}_{iV(\mathbf{Y}_{ik}) k } - \boldsymbol{\mu}_{V(\mathbf{Y}_{ik})k}\right)^\top \Big(\boldsymbol{\Sigma}_{V(\mathbf{Y}_{ik})}\Big)^{-1}\left(\Tilde{\X}_{i V(\mathbf{Y}_{ik}) k } - \boldsymbol{\mu}_{V(\mathbf{Y}_{ik})k}\right)\Bigg] \ .
	\end{split}    
\end{equation*}
When $p > n$ the usual Maximum Likelihood Estimator (MLE) of $\boldsymbol{\pi}_k$, $\boldsymbol{\mu}_k$ and $\boldsymbol{\Sigma}$ is not well defined, since the sample variance and covariance matrix is not positive definite. We regularize the problem through a Ridge-like penalty. Specifically, the regularized log--likelihood $\ell^{\boldsymbol{\lambda}}(\boldsymbol{\theta})$ can be written as
\begin{equation} \label{eq:loglik:lambda}
	\ell^{\boldsymbol{\lambda}}(\boldsymbol{\pi}, \boldsymbol{\mu}, \boldsymbol{\Sigma}) = \ell(\boldsymbol{\pi}, \boldsymbol{\mu}, \boldsymbol{\Sigma}) - \frac{1}{2} P(\boldsymbol{\lambda}, \boldsymbol{\Sigma}) ,
\end{equation}

where the penalization term $P(\boldsymbol{\lambda}, \boldsymbol{\Sigma})$ is defined as 
\begin{equation}
	P(\boldsymbol{\lambda}, \boldsymbol{\Sigma}) := \sum_{k = 1}^K \sum_{i=1}^{n_k} \text{tr}\left(\boldsymbol{\Lambda}_{V(\Y_{ik})} \boldsymbol{\Sigma}_{V(\Y_{ik})}^{-1}\right),
\end{equation}
with $\boldsymbol{\lambda} = (\lambda_1, \lambda_2, \ldots, \lambda_p)^\top \in \R_+^p$ and $\boldsymbol{\Lambda} = \operatorname{diag}(\boldsymbol{\lambda})$. In the low-dimensional setting with $n > p$ one can set $\boldsymbol{\lambda}=\boldsymbol{0}$. The penalized Maximum Likelihood Estimators of the parameters involved in the regularized likelihood \eqref{eq:loglik:lambda} can be obtained in closed form (see details in Appendix \ref{appendixA}) and are given by
\begin{align}                
	\label{eq:estimators p>n pi}
	\hat{\pi}_k(s) &= \frac{s!(p-s)!}{p!}  \frac{\sum_{i=1}^{n_k}  \ind\left(|V(\Y_{ik})|= s\right) }{n_k},\\
	\label{eq:estimators p>n mu}
	\hat{\mu}_{jk} &= \frac{\sum_{i=1}^{n_k} \tilde{X}_{ijk} Y_{ijk}}{\sum_{i=1}^{n_k} Y_{ijk} },\\
	\label{eq:estimators p>n Sigma}
	\hat{\boldsymbol{\Sigma}}^{\boldsymbol{\lambda}} &= \frac{\sum_{k = 1}^K \sum_{i = 1}^{n_k} \left[
		\Y_{ik}\Y_{ik}^\top              
		\circ
		\left(\left(
		\tilde{\X}_{ik} - \hat{\boldsymbol{\mu}}_k
		\right)\left(
		\tilde{\X}_{ik} - \hat{\boldsymbol{\mu}}_k
		\right)^\top + \Lambda \right)
		\right]}{\sum_{k = 1}^K \sum_{i = 1}^{n_k} \Y_{ik}\Y_{ik}^\top},
\end{align}
with $ j = 1, \ldots, p$ and $k = 1, \ldots, K$. Note that the fraction in the variance and covariance matrix estimator is considered to be entry-wise and $\circ$ denotes the Hadamard product. When $\boldsymbol{\lambda}=\boldsymbol{0}$ the expression above coincides with the MLE\citep{farcomeni2016manova}.
A general requirement is that $\sum_{k = 1}^K \sum_{i = 1}^{n_k} Y_{ij_1k}Y_{ij_2k} > 0$ for each $j_1 \neq j_2, \; j_1, j_2 = 1, \dots, p$, i.e. each couple of variables is simultaneously present in at least one observation in the sample. In real applications with $p >> n$ there might be couples of variables that do not satisfy the previous condition. Those variables must be removed from the data set starting from those with the largest number of absent components, until all the others satisfy the condition. This process reduces the dimensionality, but it has an impact only on the continuous part of the likelihood. 

\subsection{Hypothesis testing}
\label{sec:test}

In this section we define a MANOVA test in our possibly high--dimensional semicontinuos setting.
The null hypothesis we wish to test, under the framework previously described, can be summarized as
\begin{equation*}
	H_0: \left(\bigcap_{s=0}^p \pi_{1}(s)= \dots = \pi_{K}(s)\right) \cap \left(\bigcap_{j=1}^p \mu_{j1}=\dots = \mu_{jK}\right),
\end{equation*}
which corresponds to assuming the probability of having $s$ positive components, with $s = 0, \dots, p$, is equal in all the groups, and, given a presence, all expectations are the same.
We let $\Theta_0 \subseteq \Theta$ be the parametric space under the null hypothesis. For the sake of generality we admit the possibility of a penalty parameter
$\boldsymbol{\lambda}_0 = (\lambda_{01}, \lambda_{02}, \ldots, \lambda_{0p})^\top$ under the null hypothesis that can differ from $\boldsymbol{\lambda}$, the penalty parameter under the alternative hypothesis. The log--likelihood under $H_0$ can be written as
\begin{equation*}
	\begin{split}
		\ell_0\left(\boldsymbol{\pi}_0, \boldsymbol{\mu}_0, \boldsymbol{\Sigma}_0\right)
		&= \sum_{k=1}^K \sum_{i=1}^{n_k} \sum_{\boldsymbol{a}} \ind(V(\boldsymbol{a})=V(\mathbf{Y}_{ik}))
		\log(\pi_{0}(\boldsymbol{a})) + \sum_{k=1}^K \sum_{i=1}^{n_k} \Bigg[ \frac{-n_{ik}}{2}\log(2\pi)+\\
		&-\frac{1}{2}\log\left(\left|\boldsymbol{\Sigma}_{V(\mathbf{Y}_{ik})0} \right|\right)+\\
		&-\frac{1}{2} \left(\Tilde{\X}_{iV(\mathbf{Y}_{ik}) k } - \boldsymbol{\mu}_{V(\mathbf{Y}_{ik})0}\right)^\top \Big(\boldsymbol{\Sigma}_{V(\mathbf{Y}_{ik})0}\Big)^{-1}\left(\Tilde{\X}_{i V(\mathbf{Y}_{ik}) k } - \boldsymbol{\mu}_{V(\mathbf{Y}_{ik})0}\right)\Bigg] \\
	\end{split}
\end{equation*}
and its penalized version is then
\begin{equation} \label{eq:loglik:lambda:H0}
	\ell_0^{\boldsymbol{\lambda}_0}\left(\boldsymbol{\pi}_0, \boldsymbol{\mu}_0, \boldsymbol{\Sigma}_0\right) = \ell_0\left(\boldsymbol{\pi}_0, \boldsymbol{\mu}_0, \boldsymbol{\Sigma}_0\right) -\frac{1}{2} P(\boldsymbol{\lambda}_0, \boldsymbol{\Sigma}_0).\\
\end{equation}
It can be shown, similarly to the unconstrained case, that the expression above is maximized by 
\begin{align}
	\label{eq:estimators p>n H0 pi}
	\hat{\pi}_0(s) &= \frac{s!(p-s)!}{p!}  \frac{\sum_{k=1}^K \sum_{i=1}^{n_k}  \ind\left(|V(\Y_{ik})|= s\right) }{n},\\
	\label{eq:estimators p>n H0 mu}
	\hat{\mu}_{j0} &= \frac{\sum_{k = 1}^K \sum_{i = 1}^{n_k} \tilde{X}_{ijk}Y_{ijk}}{\sum_{k = 1}^K \sum_{i = 1}^{n_k} Y_{ijk}},\\
	\label{eq:estimators p>n H0 Sigma}
	\hat{\boldsymbol{\Sigma}}_0^{\boldsymbol{\lambda}_0} &= \frac{\sum_{k = 1}^K \sum_{i = 1}^{n_k} \left[
		\Y_{ik}\Y_{ik}^\top 
		\circ
		\left(\left(
		\tilde{\X}_{ik} - \hat{\boldsymbol{\mu}}_0
		\right)\left(
		\tilde{\X}_{ik} - \hat{\boldsymbol{\mu}}_0
		\right)^\top + \Lambda_0 \right)
		\right]}{\sum_{k = 1}^K \sum_{i = 1}^{n_k} \Y_{ik}\Y_{ik}^\top}. 
\end{align}
where $\Lambda_0 = \operatorname{diag}(\boldsymbol{\lambda}_0)$.
In order to test $H_0$ we define a Likelihood Ratio Test (LRT) statistic, which can be obtained from the regularized likelihoods as

\begin{equation}
	\label{eq:test statistic}
	\begin{split}
		D^{\boldsymbol{\lambda}, \boldsymbol{\lambda}_0} & = -2 \log\left(\frac{\sup_{\boldsymbol{\theta} \in \Theta} \mathcal{L}(\boldsymbol{\theta})}{\sup_{\boldsymbol{\theta} \in \Theta_0}\mathcal{L}_0(\boldsymbol{\theta})} \right) \\
		& = -2 \left(\ell(\hat{\boldsymbol{\pi}}, \hat{\boldsymbol{\mu}}, \hat{\boldsymbol{\Sigma}}^{\boldsymbol{\lambda}}) - \ell_0(\hat{\boldsymbol{\pi}}_0, \hat{\boldsymbol{\mu}}_0, \hat{\boldsymbol{\Sigma}}_0^{\boldsymbol{\lambda}_0})\right),
	\end{split}
\end{equation}
where $\mathcal{L} = \exp(\ell)$ and $\mathcal{L}_0 = \exp(\ell_0)$ are the likelihoods. When $n>p$ and $\boldsymbol{\lambda} = \boldsymbol{\lambda}_0 = 0$ the expression above reduces to the
classical LRT statistic, which Wilks' theorem \citep{wilks} guarantees to
be asymptotically distributed like a chi-squared, with degrees of freedom given by the difference in the number of free parameters. This result is not bound to hold in general, therefore we rely on a permutation test \citep{pesa:01,miel:berr:07}, based on the test statistic \eqref{eq:test statistic}.

The choice of the values of $\boldsymbol{\lambda}$ and $\boldsymbol{\lambda}_0$ used in $ D^{\boldsymbol{\lambda}, \boldsymbol{\lambda}_0 }$ is discussed in the next section \ref{subsec:Choice_lambda}.

\subsection{Choice of $\lambda$ and $\lambda_0$}
\label{subsec:Choice_lambda}

The selection of $\boldsymbol{\lambda}$ and $\boldsymbol{\lambda}_0$ can be performed by Cross--Validation or by information criteria
\begin{equation*}
	IC(\lambda, \alpha) = \text{Goodness of the fit} + \text{Penalty} \times \text{Model complexity measure}.
\end{equation*}
The second method is faster and what we suggest to use in practice.
Measures of model complexity have been extensively studied \citep{janson}.
We use the trace of the Fisher information matrix \citep{takeuchi,bozdogan}, as we detail below. 

Let $\Phi$ and $\Phi_0$ denote the sets of $\boldsymbol{\lambda}$ and $\boldsymbol{\lambda}_0$ that guarantee the positive definiteness of the variance and covariance matrix estimators \eqref{eq:estimators p>n Sigma} and \eqref{eq:estimators p>n H0 Sigma}, respectively.
The choice of $\boldsymbol{\lambda} \in \Phi$ is based on the minimization of
\begin{align}      
	\label{eq:ourAIC}
	\hat{\boldsymbol{\lambda}} &= \arg\min_{\boldsymbol{\lambda} \in \Phi} M(\boldsymbol{\lambda}, \hat{\boldsymbol{\pi}}, \hat{\boldsymbol{\mu}}, \hat{\boldsymbol{\Sigma}}) \\
	&= \arg\min_{\boldsymbol{\lambda} \in \Phi} -2 \ell\left(\hat{\boldsymbol{\pi}}, \hat{\boldsymbol{\mu}}, \hat{\boldsymbol{\Sigma}}^{\boldsymbol{\lambda}}\right) + \left(\log(n)+\frac{1}{2}\log(p)\right) \sum_{k = 1}^K \sum_{i = 1}^{n_k}\text{tr}
	\left(\left(\hat{\boldsymbol{\Sigma}}^{\boldsymbol{\lambda}}_{V(\Y_{ik})}\right)^{-1}\right).
\end{align}
The expressions $ M(\boldsymbol{\lambda}, \hat{\boldsymbol{\pi}}, \hat{\boldsymbol{\mu}}, \hat{\boldsymbol{\Sigma}})$ are obtained by reformulating our problem as a weighted regression problem, and using the trace of the Fisher information matrix as a measure of the complexity of the model \citep{bozdogan} (see Appendix \ref{appendixB} for the complete derivation).
In the same spirit we choose $\boldsymbol{\lambda}_0 \in \Phi_0$ such that
\begin{align}      
	\label{eq:ourAIC_H0}
	\hat{\boldsymbol{\lambda}}_0 &= \arg\min_{\boldsymbol{\lambda}_0 \in \Phi_0} M(\boldsymbol{\lambda}_0, \hat{\boldsymbol{\pi}}_0, \hat{\boldsymbol{\mu}}_0, \hat{\boldsymbol{\Sigma}}_0) \\
	&= \arg\min_{\boldsymbol{\lambda}_0 \in \Phi_0} -2 \ell_0\left(\hat{\boldsymbol{\pi}}_0, \hat{\boldsymbol{\mu}}_0, \hat{\boldsymbol{\Sigma}}_0^{\boldsymbol{\lambda}_0}\right) + \left(\log(n)+\frac{1}{2}\log(p)\right) \sum_{k = 1}^K \sum_{i = 1}^{n_k}\text{tr}
	\left(\left(\hat{\boldsymbol{\Sigma}}^{\boldsymbol{\lambda}_0}_{0, V(\Y_{ik})}\right)^{-1}\right).
\end{align}
The LRT test statistic \eqref{eq:test statistic} is finally denoted as $D^{\hat{\boldsymbol{\lambda}}, \hat{\boldsymbol{\lambda}}_0}$.

\section{Simulation Study}
\label{sec:sim}

In this section we present a simulation study accomplished to evaluate the performance of the new test.
We consider $K = 2$ and $K = 4$ groups, with balanced sample sizes $n_1 = \dots = n_K$, with different number of observations in each group, $n_k = \{5, 10\}$, and different dimensionality, $p = \{50, 100, 150, 200\}$. For the variance and covariance matrix we assume the logarithms of the positive observations have unit variance and constant covariance, equal to $\rho$, with $\rho = \{0, 0.4\}$. We set the mean vector of the first group $\boldsymbol{\mu}_1$ equal to the null vector, while the other groups have mean components $\mu_{jk} = \mu_{j1} + c_1 (k - 1)/(K - 1)$, with $c_1 = \{0, 1, 5\}$, for $j = 1, \dots, p$. The marginal probability of having zeros in the first group is varied as $\pi_{j1} = \{0.2, 0.5, 0.8\}$ for $j = 1, \dots, p$, while for the $k$-th group $\pi_{jk} = \pi_{j1} + c_{2}(k - 1)/(K - 1)$, where $c_2 = \{0, 0.15, 0.3\}$, whenever $\pi_{jk} < 1$. 

Considering all the parameters' combinations we finally evaluate 448 scenarios. For each scenario we generate data 1000 times, we fix a nominal test size at $\alpha=0.05$, and use 1000 permutations to evaluate the significance level of our proposed test statistic. During the estimation procedure, we additionally assume $\lambda := \lambda_1 = \dots = \lambda_p$ and $\lambda_0 = \lambda_{01} = \dots = \lambda_{0p}$.

Table \ref{Tab::PropRejectionSimK2-a}, \ref{Tab::PropRejectionSimK2-b} and \ref{Tab::PropRejectionSimK2-c} show the proportions of rejections averaged over the $1000$ replicates for each scenario with $K = 2$, while Table \ref{Tab::PropRejectionSimK4} presents the rejections when $K = 4$. When testing the null hypothesis with two groups, we compare our results to those of \cite{chen2011regularized}, as it is the only approach in the literature capable of handling high-dimensional data in the presence of missing data. However, this method is limited to two-group scenarios, meaning the comparison cannot be extended to cases with $K = 4$ groups.
Note that for all scenarios in which $c_1=c_2=0$ the null hypothesis is true, and the proportion of rejections corresponds to the observed test level. In all other cases, the proportion of rejections corresponds to the observed power. The actual size of the regularized MANOVA test closely aligns with the nominal size of $0.05$.
In some instances, the proportion of rejections slightly exceeds $0.05$, but this excess constantly remains below the Monte Carlo error and could be reduced by increasing the number of repetitions. On the other hand, the test proposed by \cite{chen2011regularized} significantly exceeds the nominal size when the number of observations is small ($n_k = 5$) and the proportion of missing data is high ($\pi_{j1} = 0.8$). This could be because the method was not primarily developed for hypothesis testing in the presence of missing data. When a large number of components are missing and insufficient information can be recovered from the remaining observations in the sample, it faces difficulties in reaching the correct conclusion.
As could be expected the power of the regularized MANOVA test increases with the sample size and with $c_1$ and $c_2$ growth, which leads to more separation between the two groups. With respect to \cite{chen2011regularized}, in the configurations in which the nominal size was garanteed, our method obtains in general higher or comparable power.
The effect of positive correlation among the components leads in the new proposed method to a small decrease in the proportion of rejections. In the cases in which $c_2 =0$, we additionally observe that higher values of $\pi_{j1}$ lead to lower power. This can be explained by the fact some variables are removed when there is a high probability of zero measurements.
Indeed, as described in Subsection \ref{subsec:likelihood inference}, some components can be removed during the estimation procedure, resulting on estimates of dimension $p^*$, with $p^* \leq p$. Table \ref{Tab::DimLambdaSIM-K2} and \ref{Tab::DimLambdaSIM-K4} in Appendix \ref{appendixC} report for each scenario the mean over $1000$ replicates of $p^*$ and of the selected values of $\lambda$ and $\lambda_0$ for $K = 2$ and $K = 4$ respectively. The mean value of $p^*$ is notably influenced by the marginal probabilities of missing data in the two groups $\pi_{j1}$ and $\pi_{j2}$. Consequently, we observe a significant decrease in $p^*$ when $\pi_{j1}$ and/or $c_2$ grow. As expected, when the number of removed variables is too large, the observed power drops significantly. For instance, in the scenario with $(K, n_k, c_1, c_2, \pi_{j1}, \rho) = (4, 5, 0, 0.15, 0.8, 0)$ the rejection rate is extremely low. This can be attributed to the very small value of $p^*$ in this setting. On the other hand, when the number of observations slightly grows ($n_k = 10$), our method is able to recover a satisfactory power. Therefore, it can be concluded that, when $p^*$ is sufficiently large, the regularized MANOVA test demonstrates good power.

We simulate also two additional scenarios that mimic the real example described in Subsection \ref{subsec:microRNA}. Hence we test a situation with $n_1 = n_2 = 5$, $p = 339$, both under the null (scenario A) and under the alternative hypothesis (scenario B). The mean of the normal distribution used to simulate the data in scenarios A and B is equal to the sample mean of the blastocyst data presented in Subsection \ref{subsec:microRNA} under the null and under the alternative hypothesis respectively. In the second case, some of the components of the sample means of the different groups were missing, therefore we imputed the missing values with the corresponding components under $H_0$. In a similar vein, the marginal probability of a missing component is equal to its sample version in the two different scenarios. Also the sample variance and covariance matrixes had some missing entrances. Hence we have imputed the missing elements along the main diagonal with $1$ and the ones that were not on the main diagonal with $0$. To achieve the positive definiteness, we additionally added $0.1$ to each element on the main diagonal. The proportion of rejections in scenario A is equal to $0.051$, while under scenario B it is $0.365$. The mean dimension of components used for the estimation is respectively $251.74$ and $170.64$.

\begin{table}[h!]
	\centering
\begin{adjustbox}{max width=1\textwidth} 
	\begin{tabular}{ccccc|cccc|cccc}
				& & & & &\multicolumn{4}{c|}{$n_k = 5$} & \multicolumn{4}{c}{$n_k = 10$}\\
				$c_1$ & $c_2$ & $\pi_{j1}$ & $\rho$ & test & \multicolumn{1}{c}{$p = 50$} & \multicolumn{1}{c}{$p = 100$} & \multicolumn{1}{c}{$p = 150$} & \multicolumn{1}{c|}{$p = 200$} & \multicolumn{1}{c}{$p = 50$} & \multicolumn{1}{c}{$p = 100$} & \multicolumn{1}{c}{$p = 150$} & \multicolumn{1}{c}{$p = 200$}\\ 
				\hline
				0 & 0 & 0.2 & 0 & scMAN & 0.045 & 0.038 & 0.058 & 0.055 & 0.041 & 0.058 & 0.062 & 0.046 \\ 
				&  &  &  & Chen & 0.045 & 0.046 & 0.045 & 0.055 & 0.035 & 0.049 & 0.052 & 0.058 \\
				0 & 0 & 0.2 & 0.4 & scMAN & 0.048 & 0.048 & 0.048 & 0.053 & 0.044 & 0.055 & 0.043 & 0.058 \\
				&  &  &  & Chen & 0.051 & 0.042 & 0.046 & 0.064 & 0.040 & 0.045 & 0.052 & 0.049 \\
				0 & 0 & 0.5 & 0 & scMAN & 0.046 & 0.051 & 0.038 & 0.043 & 0.031 & 0.053 & 0.047 & 0.052 \\
				&  &  &  & Chen & 0.049 & 0.079 & 0.033 & 0.051 & 0.316 & 0.080 & 0.051 & 0.051 \\
				0 & 0 & 0.5 & 0.4 & scMAN & 0.039 & 0.043 & 0.040 & 0.043 & 0.042 & 0.050 & 0.045 & 0.053 \\ 
				&  &  &  & Chen & 0.050 & 0.056 & 0.040 & 0.056 & 0.312 & 0.078 & 0.049 & 0.047 \\
				0 & 0 & 0.8 & 0 & scMAN & 0.050 & 0.051 & 0.045 & 0.040 & 0.048 & 0.060 & 0.057 & 0.050 \\
				&  &  &  & Chen & 0.756 & 0.536 & 0.376 & 0.245 & 1.000 & 1.000 & 1.000 & 1.000 \\
				0 & 0 & 0.8 & 0.4 & scMAN & 0.046 & 0.048 & 0.047 & 0.031 & 0.046 & 0.055 & 0.055 & 0.047 \\
				&  &  &  & Chen & 0.757 & 0.533 & 0.380 & 0.251 & 1.000 & 1.000 & 1.000 & 1.000 \\
				\hline
			\end{tabular}
    \end{adjustbox}
	\caption{Observed test level for the regularized MANOVA test (scMAN) and for \cite{chen2011regularized} (Chen) in presence of semicontinuous high-dimensional data for $K = 2$ and different values of $n$, $p$, $\pi_{j1}$, $\rho$. The nominal size is $0.05$ and the presented results are the average of $1000$ repetitions.}
	\label{Tab::PropRejectionSimK2-a}
\end{table}

\begin{table}[h!]
	\centering
\begin{adjustbox}{max width=1\textwidth} 
	\begin{tabular}{ccccc|cccc|cccc}
				& & & & &\multicolumn{4}{c|}{$n_k = 5$} & \multicolumn{4}{c}{$n_k = 10$}\\
				$c_1$ & $c_2$ & $\pi_{j1}$ & $\rho$ & test & \multicolumn{1}{c}{$p = 50$} & \multicolumn{1}{c}{$p = 100$} & \multicolumn{1}{c}{$p = 150$} & \multicolumn{1}{c|}{$p = 200$} & \multicolumn{1}{c}{$p = 50$} & \multicolumn{1}{c}{$p = 100$} & \multicolumn{1}{c}{$p = 150$} & \multicolumn{1}{c}{$p = 200$}\\ 
				\hline
				1 & 0 & 0.2 & 0 & scMAN & 0.985 & 0.994 & 0.990 & 0.984 & 1.000 & 1.000 & 1.000 & 1.000 \\ 
				&  &  &  & Chen & 0.851 & 0.989 & 0.998 & 1.000 & 0.594 & 0.824 & 0.935 & 0.967 \\
				1 & 0 & 0.2 & 0.4 & scMAN & 0.505 & 0.550 & 0.529 & 0.531 & 0.736 & 0.837 & 0.876 & 0.891 \\
				&  &  &  & Chen & 0.402 & 0.481 & 0.494 & 0.509 & 0.382 & 0.549 & 0.644 & 0.702 \\
				1 & 0 & 0.5 & 0 & scMAN & 0.755 & 0.820 & 0.714 & 0.725 & 0.988 & 0.959 & 0.939 & 0.916 \\
				&  &  &  & Chen & 0.205 & 0.268 & 0.314 & 0.326 & 0.323 & 0.108 & 0.104 & 0.133 \\
				1 & 0 & 0.5 & 0.4 & scMAN & 0.420 & 0.469 & 0.451 & 0.443 & 0.770 & 0.767 & 0.769 & 0.794 \\
				&  &  &  & Chen & 0.203 & 0.248 & 0.295 & 0.268 & 0.326 & 0.114 & 0.109 & 0.131 \\ 
				1 & 0 & 0.8 & 0 & scMAN & 0.134 & 0.186 & 0.237 & 0.219 & 0.243 & 0.406 & 0.548 & 0.634 \\
				&  &  &  & Chen & 0.758 & 0.534 & 0.392 & 0.249 & 1.000 & 1.000 & 1.000 & 1.000 \\
				1 & 0 & 0.8 & 0.4 & scMAN & 0.122 & 0.161 & 0.169 & 0.172 & 0.250 & 0.359 & 0.45 & 0.514 \\
				&  &  &  & Chen & 0.758 & 0.537 & 0.392 & 0.258 & 1.000 & 1.000 & 1.000 & 1.000 \\
				5 & 0 & 0.2 & 0 & scMAN & 1.000 & 1.000 & 1.000 & 1.000 & 1.000 & 1.000 & 1.000 & 1.000 \\
				&  &  &  & Chen & 1.000 & 1.000 & 1.000 & 1.000 & 0.994 & 1.000 & 1.000 & 1.000 \\
				5 & 0 & 0.2 & 0.4 & scMAN & 0.999 & 0.999 & 1.000 & 1.000 & 1.000 & 1.000 & 1.000 & 1.000 \\ 
				&  &  &  & Chen & 1.000 & 1.000 & 1.000 & 1.000 & 0.995 & 1.000 & 1.000 & 1.000 \\
				5 & 0 & 0.5 & 0 & scMAN & 0.999 & 1.000 & 0.999 & 1.000 & 1.000 & 1.000 & 1.000 & 1.000 \\
				&  &  &  & Chen & 0.767 & 0.837 & 0.929 & 0.963 & 0.327 & 0.123 & 0.138 & 0.199 \\
				5 & 0 & 0.5 & 0.4 & scMAN & 0.999 & 1.000 & 0.999 & 0.988 & 1.000 & 1.000 & 1.000 & 1.000 \\
				&  &  &  & Chen & 0.775 & 0.857 & 0.944 & 0.964 & 0.327 & 0.123 & 0.138 & 0.199 \\
				5 & 0 & 0.8 & 0 & scMAN & 0.688 & 0.584 & 0.550 & 0.521 & 0.976 & 0.976 & 0.987 & 0.977 \\
				&  &  &  & Chen & 0.755 & 0.540 & 0.401 & 0.263 & 1.000 & 1.000 & 1.000 & 1.000 \\
				5 & 0 & 0.8 & 0.4 & scMAN & 0.639 & 0.556 & 0.543 & 0.482 & 0.972 & 0.978 & 0.991 & 0.985 \\ 
				&  &  &  & Chen & 0.755 & 0.540 & 0.398 & 0.264 & 1.000 & 1.000 & 1.000 & 1.000 \\
				\hline
			\end{tabular}
\end{adjustbox}
	\caption{Observed power for the regularized MANOVA test (scMAN) and for \cite{chen2011regularized} (Chen) in presence of semicontinuous high-dimensional data for $K = 2$ and different values of $n$, $p$, $c_1$, $\pi_{j1}$, $\rho$. The nominal size is $0.05$ and the presented results are the average of $1000$ repetitions.}
	\label{Tab::PropRejectionSimK2-b}
\end{table}

\begin{table}[h!]
	\centering
\begin{adjustbox}{max width=1\textwidth} 
	\begin{tabular}{ccccc|cccc|cccc}
				& & & & &\multicolumn{4}{c|}{$n_k = 5$} & \multicolumn{4}{c}{$n_k = 10$}\\
				$c_1$ & $c_2$ & $\pi_{j1}$ & $\rho$ & test & \multicolumn{1}{c}{$p = 50$} & \multicolumn{1}{c}{$p = 100$} & \multicolumn{1}{c}{$p = 150$} & \multicolumn{1}{c|}{$p = 200$} & \multicolumn{1}{c}{$p = 50$} & \multicolumn{1}{c}{$p = 100$} & \multicolumn{1}{c}{$p = 150$} & \multicolumn{1}{c}{$p = 200$}\\ 
				\hline 
				0 & 0.15 & 0.2 & 0 & scMAN & 0.157 & 0.132 & 0.089 & 0.060 & 0.546 & 0.576 & 0.502 & 0.426 \\
				&  &  &  & Chen & 0.079 & 0.110 & 0.149 & 0.164 & 0.139 & 0.215 & 0.278 & 0.334 \\
				0 & 0.15 & 0.2 & 0.4 & scMAN & 0.107 & 0.075 & 0.056 & 0.059 & 0.455 & 0.308 & 0.168 & 0.117 \\
				&  &  &  & Chen & 0.073 & 0.079 & 0.089 & 0.089 & 0.129 & 0.174 & 0.197 & 0.210 \\
				0 & 0.15 & 0.5 & 0 & scMAN & 0.179 & 0.100 & 0.051 & 0.016 & 0.360 & 0.270 & 0.146 & 0.084 \\
				&  &  &  & Chen & 0.093 & 0.096 & 0.099 & 0.125 & 0.914 & 0.755 & 0.589 & 0.445 \\ 
				0 & 0.15 & 0.5 & 0.4 & scMAN & 0.149 & 0.107 & 0.060 & 0.043 & 0.284 & 0.157 & 0.078 & 0.064 \\
				&  &  &  & Chen & 0.089 & 0.097 & 0.102 & 0.121 & 0.912 & 0.755 & 0.594 & 0.438 \\
				0 & 0.15 & 0.8 & 0 & scMAN & 0.269 & 0.379 & 0.426 & 0.474 & 0.936 & 0.948 & 0.838 & 0.671 \\
				&  &  &  & Chen & 0.994 & 0.985 & 0.978 & 0.965 & 1.000 & 1.000 & 1.000 & 1.000 \\
				0 & 0.15 & 0.8 & 0.4 & scMAN & 0.261 & 0.403 & 0.438 & 0.473 & 0.926 & 0.896 & 0.778 & 0.648 \\
				&  &  &  & Chen & 0.994 & 0.986 & 0.977 & 0.967 & 1.000 & 1.000 & 1.000 & 1.000 \\ 
				0 & 0.3 & 0.2 & 0 & scMAN & 0.259 & 0.139 & 0.077 & 0.032 & 0.853 & 0.655 & 0.488 & 0.414 \\
				&  &  &  & Chen & 0.304 & 0.539 & 0.659 & 0.777 & 0.490 & 0.714 & 0.842 & 0.917 \\
				0 & 0.3 & 0.2 & 0.4 & scMAN & 0.170 & 0.096 & 0.062 & 0.050 & 0.717 & 0.317 & 0.146 & 0.101 \\
				&  &  &  & Chen & 0.235 & 0.282 & 0.331 & 0.373 & 0.450 & 0.648 & 0.746 & 0.807 \\
				0 & 0.3 & 0.5 & 0 & scMAN & 0.556 & 0.303 & 0.075 & 0.002 & 0.824 & 0.543 & 0.260 & 0.148 \\
				&  &  &  & Chen & 0.273 & 0.341 & 0.375 & 0.438 & 1.000 & 0.995 & 0.991 & 0.988 \\
				0 & 0.3 & 0.5 & 0.4 & scMAN & 0.491 & 0.330 & 0.142 & 0.040 & 0.694 & 0.348 & 0.164 & 0.113 \\
				&  &  &  & Chen & 0.258 & 0.333 & 0.373 & 0.406 & 1.000 & 0.995 & 0.990 & 0.987 \\
				\hline
			\end{tabular}
\end{adjustbox}
	\caption{Observed power for the regularized MANOVA test (scMAN) and for \cite{chen2011regularized} (Chen) in presence of semicontinuous high-dimensional data for $K = 2$ and different values of $n$, $p$, $c_2$, $\pi_{j1}$, $\rho$. The nominal size is $0.05$ and the presented results are the average of $1000$ repetitions.}
	\label{Tab::PropRejectionSimK2-c}
\end{table}

\begin{table}[h!]
	\centering
\begin{adjustbox}{max width=1\textwidth} 
			\begin{tabular}{cccc|cccc|cccc}
				& & & &\multicolumn{4}{c|}{$n_k = 5$} & \multicolumn{4}{c}{$n_k = 10$}\\
				$c_1$ & $c_2$ & $\pi_{j1}$ & $\rho$ & \multicolumn{1}{c}{$p = 50$} & \multicolumn{1}{c}{$p = 100$} & \multicolumn{1}{c}{$p = 150$} & \multicolumn{1}{c|}{$p = 200$} & \multicolumn{1}{c}{$p = 50$} & \multicolumn{1}{c}{$p = 100$} & \multicolumn{1}{c}{$p = 150$} & \multicolumn{1}{c}{$p = 200$}\\ 
				\hline
				0 & 0 & 0.2 & 0 & 0.062 & 0.067 & 0.066 & 0.066 & 0.050 & 0.042 & 0.046 & 0.053 \\
				0 & 0 & 0.2 & 0.4 & 0.056 & 0.055 & 0.057 & 0.055 & 0.048 & 0.040 & 0.061 & 0.053 \\
				0 & 0 & 0.5 & 0 & 0.041 & 0.050 & 0.054 & 0.050 & 0.060 & 0.055 & 0.049 & 0.058 \\
				0 & 0 & 0.5 & 0.4 & 0.045 & 0.037 & 0.048 & 0.039 & 0.067 & 0.055 & 0.055 & 0.059 \\
				0 & 0 & 0.8 & 0 & 0.036 & 0.042 & 0.048 & 0.065 & 0.053 & 0.053 & 0.055 & 0.061 \\
				0 & 0 & 0.8 & 0.4 & 0.037 & 0.048 & 0.042 & 0.058 & 0.046 & 0.049 & 0.049 & 0.047 \\ 
				1 & 0 & 0.2 & 0 & 0.845 & 0.991 & 1.000 & 1.000 & 0.998 & 1.000 & 1.000 & 1.000 \\
				1 & 0 & 0.2 & 0.4 & 0.349 & 0.408 & 0.444 & 0.481 & 0.546 & 0.692 & 0.760 & 0.770 \\
				1 & 0 & 0.5 & 0 & 0.487 & 0.781 & 0.846 & 0.860 & 0.941 & 1.000 & 1.000 & 1.000 \\
				1 & 0 & 0.5 & 0.4 & 0.271 & 0.334 & 0.325 & 0.356 & 0.615 & 0.764 & 0.836 & 0.816 \\
				1 & 0 & 0.8 & 0 & 0.059 & 0.084 & 0.102 & 0.127 & 0.146 & 0.247 & 0.355 & 0.436 \\
				1 & 0 & 0.8 & 0.4 & 0.056 & 0.084 & 0.086 & 0.092 & 0.147 & 0.228 & 0.281 & 0.343 \\
				5 & 0 & 0.2 & 0 & 1.000 & 0.991 & 0.984 & 0.996 & 1.000 & 1.000 & 1.000 & 1.000 \\
				5 & 0 & 0.2 & 0.4 & 1.000 & 0.992 & 0.982 & 0.998 & 1.000 & 1.000 & 1.000 & 1.000 \\ 
				5 & 0 & 0.5 & 0 & 0.881 & 0.977 & 0.996 & 1.000 & 0.993 & 1.000 & 1.000 & 1.000 \\
				5 & 0 & 0.5 & 0.4 & 0.848 & 0.946 & 0.976 & 0.988 & 0.992 & 1.000 & 0.999 & 1.000 \\
				5 & 0 & 0.8 & 0 & 0.246 & 0.340 & 0.361 & 0.374 & 0.894 & 0.952 & 0.985 & 0.984 \\
				5 & 0 & 0.8 & 0.4 & 0.225 & 0.308 & 0.329 & 0.333 & 0.857 & 0.925 & 0.965 & 0.967 \\
				0 & 0.15 & 0.2 & 0 & 0.157 & 0.224 & 0.215 & 0.225 & 0.400 & 0.486 & 0.509 & 0.526 \\
				0 & 0.15 & 0.2 & 0.4 & 0.129 & 0.115 & 0.122 & 0.088 & 0.357 & 0.370 & 0.339 & 0.248 \\
				0 & 0.15 & 0.5 & 0 & 0.129 & 0.153 & 0.138 & 0.141 & 0.320 & 0.394 & 0.341 & 0.317 \\
				0 & 0.15 & 0.5 & 0.4 & 0.115 & 0.095 & 0.08 & 0.067 & 0.274 & 0.237 & 0.205 & 0.134 \\
				0 & 0.15 & 0.8 & 0 & 0.000 & 0.006 & 0.014 & 0.040 & 0.712 & 0.901 & 0.903 & 0.822 \\
				0 & 0.15 & 0.8 & 0.4 & 0.000 & 0.006 & 0.019 & 0.036 & 0.685 & 0.859 & 0.843 & 0.763 \\
				0 & 0.3 & 0.2 & 0 & 0.343 & 0.438 & 0.460 & 0.486 & 0.852 & 0.867 & 0.818 & 0.796 \\
				0 & 0.3 & 0.2 & 0.4 & 0.266 & 0.217 & 0.184 & 0.156 & 0.814 & 0.708 & 0.498 & 0.348 \\
				0 & 0.3 & 0.5 & 0 & 0.474 & 0.543 & 0.484 & 0.367 & 0.917 & 0.930 & 0.881 & 0.796 \\
				0 & 0.3 & 0.5 & 0.4 & 0.439 & 0.400 & 0.318 & 0.219 & 0.862 & 0.755 & 0.515 & 0.313 \\
				\hline
			\end{tabular}
\end{adjustbox}
	\caption{Proportion of rejections for the regularized MANOVA test for semicontinuous high-dimensional data for $K = 4$ and different values of $n$, $p$, $c_1$, $c_2$, $\pi_{j1}$, $\rho$. The nominal size is $0.05$ and the presented results are the average of $1000$ repetitions.}
	\label{Tab::PropRejectionSimK4}
\end{table}

\section{Real Data Analyses}
\label{realdata}

We describe in this section the analysis of two real data examples. 

\subsection{Micro RNA differential expression in blastocyst cultures}
\label{subsec:microRNA}

Study of micro RNA expression is particularly useful for the assessment of
biological pathways, especially in early stage development \citep{capa:ubal:cima:noli:khal:farc:ilic:rien:16}.
In our original data example we have $n=10$ samples coming from human blastocysts (day five after {\it in vitro} fertilisation).
Simplyfing the setting, it can be said that blastocysts are composed by about one-hundred-twenty cells, about one third of which differentiated into an Inner Cell Mass (ICM), and the remaining forming an outer TrophoEctoderm (TE). The ICM will then proceed to develop into a human being,
and the TE will develop into a placenta and other annexes. 

In our data we have $n_1=5$ samples from ICM and $n_2=5$ samples from TE. MicroRNA expression of $p=339$ sequences was obtained from each sample, with technical details that can be found elsewhere \citep{capa:ubal:cima:noli:khal:farc:ilic:rien:16}. MicroRNA are short sequences
that are known to coordinate the expression of pathways of genes.
Since many pathways are not yet activated at the blastocyst stage, many micro RNA sequences will not
be found in the samples, leading to zero-inflation. This is confirmed in our data, where only about 39\% of the sequences are non-zero in all samples.

The main idea is to test the biological hypothesis that even if cells are not drastically differentiated at the blastocyst stage, their micro RNA expression already is showing different pathway activation. In our framework and notation, this precisely would correspond to rejecting 
\begin{equation}
	\label{h0gene}
	H_0: \left(\bigcap_{s=0}^p \pi_{1}(s)= \pi_{2}(s)\right) \cap \left(\bigcap_{j=1}^p \mu_{j1}=\mu_{j2}\right)
\end{equation}
indicating that there exist at least one microRNA that is expressed with differential probability or level of expression when comparing inner cell
mass with trophectoderm. Due to the very large number of zeros in the data it is clear that any test assuming multivariate normality, without taking into account the
probability masses at zero, would be biased even if well calibrated for the $p>n$ setting. 

We use our proposed test to verify \eqref{h0gene}, and obtain a test statistic 
of $15.15$ and, after permutation based on 1000 replicates, the p-value is $0.008$. We then reject the null hypothesis and conclude that there is an overall difference
in microRNA expression levels when comparing ICM and TE at day five after fertilisation. Interestingly, 53.4\% of microRNA measurements are zero in the ICM,
while 35.7\% in the TE, which indicates an higher level of biological activity for the TE. Similarly, 80.8\% estimated mean expressions are larger for TE than for ICM. 

\subsection{Alien species invasion in Socotra island}
\label{subsec:alien}

Socotra is an island located in the arabic sea and administratively belonging to Yemen. Its rich biodiversity and presence of many endemic species make it a
unique environment (e.g., \cite{att:et:al:14}, \cite{ricc:et:al:20}). 
Socotra environment is endangered by many threats, including climate change,
anthropic activities, and alien species invasion.
In this section we use an original data set about the determinants of alien species invasion in the island. See \cite{terz:et:al:18} for a similar assessment related to South Africa.

Researchers explored the island to record the abundance of $p=299$ alien species in $n=103$ plots that were approximately square shaped with an area of
about 0.5ha each. The abundance measure was the widely used Importance Value (IV), which can be calculated as
$$
IV(s) = 100 \frac{NS(s)}{\sum_u NS(u)} + 100 \frac{BA(s)}{\sum_u BA(u)},
$$
where $NS(s)$ denotes the number of stems for species $s$ that were counted in the plot, and $BA(s)$ the basal area occupied by species $s$ as the sum of areas occupied at breast height by each individual stem. 
IV ranges then from zero (species is absent) to two-hundred (the plot is entirely covered only by the species).
Despite the species considered being potentially very invasive, they are (still) absent in many plots, with about 93\% of entries being equal to zero.

Plots can be classified as being limestone, alluvial, or granite. Our main question involves the assessment of association of overall abundance of invasive alien species and the geological characteristics of the plots. We conducted a comparison across the three types of plots, yielding a p-value of 0.04 and a test statistic of 139.427. As a result, we reject the null hypothesis, concluding that there is a significant overall difference in the distribution of alien species across the different plot types.

\section{Conclusions}
\label{sec:concl}

In this paper we have presented, to the best of our knowledge, the first MANOVA test for high dimensional semicontinous data. The test depends on a penalty parameter, which can be chosen in different ways. Importantly, the permutation procedure used to compute the significance level allows the user to naturally take into account additional heterogeneity arising from any data driven penalty parameter choice.
The derivation of the formal asymptotic distribution, especially for unknown $\boldsymbol{\lambda}$, would be very cumbersome \citep{han:yin:23}. A limitation of the permutation approach is, clearly, its computational complexity. We shall report however that our implementation is rather efficient, both thanks to the closed form solution for the MLE and the use of parallel computing.
A future contribution might go beyond the parametric assumptions \citep{zhan:feng:24}, maybe based on a test statistic that involves the estimated traces of the covariance matrices \citep{chen:qin:10,sriv:kubo:13,yama:hime:15, hu:et:al:17}.

\paragraph{Software}
Software in the form of \texttt{R} package {\tt semicontMANOVA} is available on CRAN. 


\appendix
\section{Estimation of parameters}
\label{appendixA}

We want to derive $\hat{\pi}_k(s)$, $\hat{\boldsymbol{\mu}}_k$ and $\hat{\boldsymbol{\Sigma}}^{\boldsymbol{\lambda}}$, with $k = 1, \dots, K$, $s= 0, \dots, p$, that maximize the regularized log-likelihood \eqref{eq:loglik:lambda}. 
\subsection{Estimation of the parameters of the discrete part}
First of all we derive estimators of the discrete part. The constraint on the sum of the Bernoulli parameters $\sum_{\boldsymbol{a}\in \{0,1\}^p} \pi_k(\boldsymbol{a}) = 1$ can be rewritten, using the homogeneity condition \eqref{eq:assumption_pi}, obtaining 
\begin{equation}
	\label{eq:constraint Bernoulli}
	\sum_{s = 0}^p \binom{p}{s} \pi_k(s) = 1 \quad \text{ for a fixed k }\in \{1, \dots, K\}, 
\end{equation}
where the binomial coefficient is the number of $\boldsymbol{a}\in\{0, 1\}^p$ with $s$ components equal to $1$. Fixing $r\not = 0$, we take derivative of the regularized log-likelihood \eqref{eq:loglik:lambda}, with respect to $\pi_k(r)$

\begin{align}
	\frac{\partial \ell^{\boldsymbol{\lambda}}(\boldsymbol{\pi}, \boldsymbol{\mu}, \boldsymbol{\Sigma}) }{\partial \pi_k(r)} &= \sum_{i=1}^{n_k}
	\frac{-\frac{p!}{r!(p-r)!}\mathbbm{1}\left(|V(\Y_{ik})|=0\right)\pi_k(r) + \mathbbm{1}\left(|V(\Y_{ik})|= r\right) \pi_k(0)}{\left(1-\sum_{s=1}^p \frac{p!}{s!(p-s)!} \pi_k(s)\right)\pi_k(r)}.
\end{align}
The expression above is equal to zero, using \eqref{eq:constraint Bernoulli}, if 
\begin{equation}
	\label{eq:pi k for p>n}
	\hat{\pi}_k(r) = \hat{\pi}_k(0)\frac{r!(p-r)!}{p!}\frac{\sum_{i=1}^{n_k} \mathbbm{1}\left(|V(\Y_{ik})|= r\right) }{\sum_{i=1}^{n_k} \mathbbm{1}\left(|V(\Y_{ik})|= 0\right) } 
\end{equation}
for each $r = 1, \dots, p$.
Using \eqref{eq:constraint Bernoulli} we also obtain 
\begin{equation*}
	\hat{\pi}_k(0) = \frac{\sum_{i=1}^{n_k} \mathbbm{1}\left(|V(\Y_{ik})|= 0\right) }{\sum_{s = 0}^p \sum_{i=1}^{n_k} \mathbbm{1}\left(|V(\Y_{ik})|= s\right) } = \frac{\sum_{i=1}^{n_k} \mathbbm{1}\left(|V(\Y_{ik})|= 0\right) }{n_k}
\end{equation*}
and we can conclude that the estimator for the discrete part is
\begin{equation}
	\label{pi finale p>n}
	\hat{\pi}_k(r) = \frac{r!(p-r)!}{p!} \frac{\sum_{i=1}^{n_k}  \mathbbm{1}\left(|V(\Y_{ik})|= r\right) }{n_k} \qquad \text{for } r = 0, \dots, p.
\end{equation}
The derivation of the estimator under the null hypothesis follows the same reasoning after defining $\pi_0(s):=\pi_1(s) = \dots = \pi_K(s)$.

\subsection{Estimation of parameters for the continuous part}
In the expression of the likelihood $\boldsymbol{\mu}_k$ appears as subsets $\mu_{V(\Y_{ik})}$, with $i = 1, \dots, n_k$.
Note that under model specification we can separate the presence/absence and continuous parts of the likelihood. 
We have 
\begin{equation}
	\frac{\partial \ell^{\boldsymbol{\lambda}}\left(\boldsymbol{\pi}, \boldsymbol{\mu}, \boldsymbol{\Sigma}\right)}{\partial \boldsymbol{\mu}_{V(\mathbf{Y}_{ik})k}} =  \Big(\boldsymbol{\Sigma}_{V(\mathbf{Y}_{ik})} \Big)^{-1}  \left(\Tilde{\X}_{i V(\mathbf{Y}_{ik}) k } - \boldsymbol{\mu}_{V(\mathbf{Y}_{ik})k}\right).
\end{equation}
The expression above is equal to zero as soon as $\hat{\boldsymbol{\mu}}_{V(\mathbf{Y}_{ik})k} = \Tilde{\X}_{iV(\mathbf{Y}_{ik})k }$, that implies
\begin{equation}
	\label{eq:equality mu}
	\left(
	\begin{array}{c}
		\hat{\mu}_{1k} Y_{i1k} \\
		\vdots\\
		\hat{\mu}_{pk} Y_{ipk} \\
	\end{array}
	\right) = \left(
	\begin{array}{c}
		\Tilde{X}_{i1k} Y_{i1k} \\
		\vdots\\
		\Tilde{X}_{ipk} Y_{ipk},
	\end{array}
	\right)
\end{equation}
where $\tilde{\X}_{ik}$ corresponds to the vector of $\R^p$ with elements
\[
\tilde{X}_{ijk}= \begin{cases}
	\log(X_{ijk}) & \text{ if } Y_{ijk} = 1,\\
	0 & \text{ if } Y_{ijk} = 0.\\
\end{cases}
\]
Equality \eqref{eq:equality mu} holds for each $i \in \{1, \dots, n_k\}$, hence we easily obtain
\begin{equation}
	\hat{\mu}_{jk} = \frac{\sum_{i=1}^{n_k} \Tilde{X}_{ijk} Y_{ijk}}{\sum_{i=1}^{n_k} Y_{ijk} } \qquad \text{for } k = 1, \dots, K; \, j = 1, \dots, p,
\end{equation}
that coincides with the MLE \citep{farcomeni2016manova}.
In the same way we can derive the estimator under the null hypothesis $\hat{\boldsymbol{\mu}}_0$, shown in equation \eqref{eq:estimators p>n H0 mu}.

We now discuss $\hat{\boldsymbol{\Sigma}}^{\boldsymbol{\lambda}}$. We simplify the notation, indicating 
$\boldsymbol{\Sigma}_V$ in place of $\boldsymbol{\Sigma}_{V(\mathbf{Y}_{ik})}$.
For fixed $i$ and $k$ we differentiate the log-likelihood with respect to $\boldsymbol{\Sigma}_V$.
Exploiting the symmetry of $\boldsymbol{\Sigma}_V$ \citep{petersen2008matrix} we obtain 
\begin{equation}
	\begin{split}
		\frac{\partial \ell^{\boldsymbol{\lambda}}(\boldsymbol{\pi}, \boldsymbol{\mu}, \boldsymbol{\Sigma})}{\partial \boldsymbol{\Sigma}_V} &= -\frac{1}{2}\left[2\boldsymbol{\Sigma}_V ^{-1} -\boldsymbol{\Sigma}_V^{-1}\circ \I_V\right]- \frac{1}{2} \bigg[-2\boldsymbol{\Sigma}_V^{-1}\left(\Tilde{\X}_{V} - \boldsymbol{\mu}_{V}\right)\left(\Tilde{\X}_{V} - \boldsymbol{\mu}_{V}\right)^\top\boldsymbol{\Sigma}_V^{-1} + \\    &+\boldsymbol{\Sigma}_V^{-1}\left(\Tilde{\X}_{V} - \boldsymbol{\mu}_{V}\right)\left(\Tilde{\X}_{V} - \boldsymbol{\mu}_{V}\right)^\top\boldsymbol{\Sigma}_V^{-1} \circ \I_V\bigg]+\\
		&-\frac{1}{2}\left[-2\boldsymbol{\Sigma}_V^{-1}\boldsymbol{\Lambda}_V\boldsymbol{\Sigma}_V^{-1}+\boldsymbol{\Sigma}_V^{-1}\boldsymbol{\Lambda}_V\boldsymbol{\Sigma}_V^{-1}\circ \I_V\right] .
	\end{split}
\end{equation}
The expression above is equal to zero when we set 
\begin{equation}
	\label{inverse sigma estimator}
	\left(\hat{\boldsymbol{\Sigma}}_V\right)^{-1} = \left(\hat{\boldsymbol{\Sigma}}_V\right)^{-1}\left(\Tilde{\X}_{V} - \boldsymbol{\mu}_{V}\right)\left(\Tilde{\X}_{V} - \boldsymbol{\mu}_{V}\right)^\top\left(\hat{\boldsymbol{\Sigma}}_V\right)^{-1} +  \left(\hat{\boldsymbol{\Sigma}}_V\right)^{-1}\boldsymbol{\Lambda}_V\left(\hat{\boldsymbol{\Sigma}}_V\right)^{-1} ,
\end{equation}
that implies 
\begin{equation}
	\hat{\boldsymbol{\Sigma}}_V^{\boldsymbol{\lambda}} = \left(\Tilde{\X}_{V} - \boldsymbol{\mu}_{V}\right)\left(\Tilde{\X}_{V} - \boldsymbol{\mu}_{V}\right)^\top +  \boldsymbol{\Lambda}_V,
\end{equation}
where we have added the apex $\boldsymbol{\lambda}$ to underline that the estimator depends on the parameter $\boldsymbol{\lambda}$.
As before we consider the full $p \times p$ matrix, obtained summing over groups and observations, i.e.
\begin{equation}
	\label{eq:sigma lambda}
	\hat{\boldsymbol{\Sigma}}^{\boldsymbol{\lambda}} \circ \sum_{k = 1}^K \sum_{i=1}^{n_k} \left(\Y_{ik}\Y_{ik}^\top\right) = \sum_{k = 1}^K \sum_{i=1}^{n_k} \left[ \left(\Y_{ik}\Y_{ik}^\top \right) \circ \left(\left(\Tilde{\X}_{ik } - \hat{\boldsymbol{\mu}}_{k}\right)\left(\Tilde{\X}_{ik } - \hat{\boldsymbol{\mu}}_{k}\right)^\top + \boldsymbol{\Lambda} \right) \right].
\end{equation}
From \eqref{eq:sigma lambda} we directly derive the final regularized estimator for the variance and covariance matrix \eqref{eq:estimators p>n Sigma}. The estimator \eqref{eq:estimators p>n H0 Sigma} under the null hypothesis is computed with similar reasoning. 

\section{Details on the selection of penalty parameters}
\label{appendixB}

Let $n_0 := 0$ and $n_{k}^* = \sum_{t = 0}^{k-1} n_t$ be the sample size considering the observations in the first $(k-1)$-th groups, with $k = 1, \dots, K$.
We define the function $h: \N^2 \rightarrow \N$ that associates to each couple $(i, k)$ a unique scalar $l$ such that $l = h(i, k) := i + n_k^*$. It is straightforward to check that $l = 1, \dots, n$ and, for fixed $k$, $n_{k}^* + 1 \leq  h(i, k) \leq n_{k+1}^*$.
Let $\boldsymbol{Z}_l := \tilde{\X}_{ik}|\Y_{ik}$ be the $p_l$-dimensional vector, where $p_l:= \sum_{j = 1}^p Y_{ijk}$ is the number of positive components of $\Y_{ik}$. In the following discussion we will use $l$ and $(i,k)$ in an interchangeable way, to denote the specific configuration of $i$ and $k$ such that $l=h(i, k)$.

The assumption of conditional Gaussian distribution \eqref{Xtilde} leads to $\boldsymbol{Z}_l \sim \mathcal{N}_{p_l}\left(\boldsymbol{\mu}_l,\boldsymbol{\Sigma}_l\right)$, with $\boldsymbol{\mu}_l := \boldsymbol{\mu}_{V(\Y_{ik})k}$ and $\boldsymbol{\Sigma}_l = \boldsymbol{\Sigma}_{V\left(\Y_{ik}\right)}$.
Equivalently we can write $\boldsymbol{Z}_l$ as 
\[
\boldsymbol{Z}_l = \I_{p_l} \boldsymbol{\mu}_l + \boldsymbol{\epsilon}_l,
\]
where $\I_{p_l}$ is a $p_l \times p_l$ identity matrix, $\boldsymbol{\epsilon}_l \sim \mathcal{N}_{p_l}(\boldsymbol{0}, \boldsymbol{\Sigma}_{l})$ and $\boldsymbol{\epsilon}_l\indep \boldsymbol{\epsilon}_r$ for $l \not = r$. Stacking $\boldsymbol{Z}_l$ together we define 
\begin{equation}
	\label{eq:weighted LR}
	\boldsymbol{Z} = \left(
	\begin{array}{c}
		\boldsymbol{Z}_1\\
		\vdots\\
		\boldsymbol{Z}_n
	\end{array}
	\right) = \X \boldsymbol{\mu} + \boldsymbol{\epsilon},
\end{equation}
where $\X$ is a $\left(\sum_{l= 1}^n p_l\right)\times\left(\sum_{l= 1}^n p_l\right)$-dimensional matrix containing the regression coefficients, i.e.
\begin{equation}
	\label{regression coefficients}
	\X = 	\left(
	\begin{array}{ccccc}
		\I_{p_1} & \boldsymbol{0} & \boldsymbol{0} & \cdots & \boldsymbol{0}\\
		\boldsymbol{0} & \I_{p_2} & \boldsymbol{0} & \ddots & \boldsymbol{0}\\
		\boldsymbol{0} & \boldsymbol{0} & \ddots & \ddots & \boldsymbol{0}\\
		\vdots & \ddots & \ddots & \ddots & \boldsymbol{0}\\
		\boldsymbol{0} & \boldsymbol{0} & \cdots & \boldsymbol{0} & \I_{p_n}
	\end{array}\right),\\
\end{equation}
while $\boldsymbol{\mu} = \left( \boldsymbol{\mu}_1, \,
\boldsymbol{\mu}_2, \,
\cdots, \,
\boldsymbol{\mu}_n \right)^\top$ and $\boldsymbol{\epsilon} = \left( \boldsymbol{\epsilon}_1, \,
\boldsymbol{\epsilon}_2, \,
\cdots, \, \boldsymbol{\epsilon}_n\right)^\top$. Consequently $\boldsymbol{\epsilon} \sim \mathcal{N}_{\sum_{l = 1}^n p_l}\left(\boldsymbol{0}, \boldsymbol{\Omega}\right)$, where 
\begin{equation}
	\boldsymbol{\Omega} = \left(
	\begin{array}{ccccccc}
		\boldsymbol{\Sigma}_{V(\Y_{11})} &  &\boldsymbol{0} & & \cdots & & \boldsymbol{0}\\
		
		& \ddots \\		
		
		\boldsymbol{0}  &  & \boldsymbol{\Sigma}_{V(\Y_{n_1 1})} &  & \ddots &  & \vdots\\
		
		& & & \ddots \\
		\vdots & & 	\ddots   &  & \boldsymbol{\Sigma}_{V(\Y_{1n_K})} & & \boldsymbol{0}  \\
		
		& & & & & \ddots\\
		
		\boldsymbol{0}&   & 	\cdots  &   &  \boldsymbol{0}  & &  \boldsymbol{\Sigma}_{V(\Y_{n_KK})}\\
	\end{array}\right).
\end{equation} 
Hence \eqref{eq:weighted LR} corresponds to rewriting the problem as a weighted linear regression, with weights equal to the inverse of the variance and covariance matrix $\boldsymbol{\Omega}$. The Fisher information matrix of the problem is \[
\mathcal{\I} = \X^\top \boldsymbol{\Omega}^{-1}\X = \boldsymbol{\Omega}^{-1},
\]
and its trace can be used as a measure of the complexity of the model \citep{takeuchi, bozdogan, boisbunon2013aic}. Since $\boldsymbol{\Omega}$ is a diagonal matrix, we have 
\begin{equation}
	\label{eq:trace Fisher}
	\text{tr}(\mathcal{\I}) =
	\text{tr}(\boldsymbol{\Omega}^{-1}) = \sum_{k = 1}^K \sum_{i = 1}^{n_k} \text{tr}\left(\boldsymbol{\Sigma}^{-1}_{V(\Y_{ik})}\right).
\end{equation}
We mimic the form of the Akaike information criterion (AIC) and define the quantity $ M(\boldsymbol{\lambda}, \hat{\boldsymbol{\pi}}, \hat{\boldsymbol{\mu}}, \hat{\boldsymbol{\Sigma}})$
\begin{equation}
	M(\boldsymbol{\lambda}, \hat{\boldsymbol{\pi}}, \hat{\boldsymbol{\mu}}, \hat{\boldsymbol{\Sigma}}):=  -2\log\left(\mathcal{L}^{\boldsymbol{\lambda}}\left(\hat{\boldsymbol{\pi}}, \hat{\boldsymbol{\mu}}, \hat{\boldsymbol{\Sigma}}^{\boldsymbol{\lambda}}\right)\right) + \left(\log(n)+\frac{1}{2}\log(p)\right)  \sum_{k = 1}^K \sum_{i = 1}^{n_k}\text{tr}
	\left(\left(\hat{\boldsymbol{\Sigma}}^{\boldsymbol{\lambda}}_{V(\Y_{ik})}\right)^{-1}\right).
\end{equation}
This quantity combines the goodness of the model and its complexity, therefore its is used to select $\hat{\boldsymbol{\lambda}}$, as shown in \eqref{eq:ourAIC}.

\section{Additional tables}
\label{appendixC}

\begin{table}[!h]
	 \rotatebox{270}{
		\begin{minipage}{1.45\textwidth}
			\centering
			\begin{adjustbox}{max width=\textwidth} 
				\begin{tabular}{cccc|cccc|cccc}
					& & & &\multicolumn{4}{c|}{$n_k = 5$} & \multicolumn{4}{c}{$n_k = 10$}\\
					$c_1$ & $c_2$ & $\pi_{j1}$ & $\rho$ & \multicolumn{1}{c}{$p = 50$} & \multicolumn{1}{c}{$p = 100$} & \multicolumn{1}{c}{$p = 150$} & \multicolumn{1}{c|}{$p = 200$} & \multicolumn{1}{c}{$p = 50$} & \multicolumn{1}{c}{$p = 100$} & \multicolumn{1}{c}{$p = 150$} & \multicolumn{1}{c}{$p = 200$}\\ 
					\hline
					0 & 0 & 0.2 & 0 & 49.93; 4.37-4.41 & 99.82; 4.78-4.84 & 149.66; 5.03-5.10 & 199.50; 5.21-5.30 & 50.00; 5.01-5.01 & 100.00; 5.41-5.42 & 150; 5.65-5.66 & 200.00; 5.82-5.83 \\ 
					0 & 0 & 0.2 & 0.4 & 49.93; 4.39-4.44 & 99.82; 4.83-4.92 & 149.66; 5.12-5.30 & 199.50; 5.46-5.78 & 50.00; 5.03-5.04 & 10.000; 5.43-5.44 & 150; 5.67-5.68 & 200.00; 5.84-5.86 \\ 
					0 & 0 & 0.5 & 0 & 37.22; 4.36-4.59 & 69.50; 4.96-5.53 & 100.75; 5.49-6.42 & 130.81; 6.10-7.45 & 47.90; 5.16-5.23 & 93.54; 5.75-5.92 & 137.87; 6.13-6.4 & 181.67; 6.5-6.93 \\ 
					0 & 0 & 0.5 & 0.4 & 37.22; 4.46-4.84 & 69.50; 5.22-6.12 & 100.75; 6.07-7.61 & 130.81; 7.09-9.13 & 47.90; 5.26-5.39 & 93.54; 5.98-6.41 & 137.87; 6.74-7.49 & 181.67; 7.63-8.71 \\ 
					0 & 0 & 0.8 & 0 & 7.99; 3.35-3.42 & 13.61; 3.66-3.82 & 19.09; 3.87-4.15 & 24.26; 4.02-4.41 & 13.38; 4.38-4.48 & 21.47; 4.70-4.92 & 27.68; 4.92-5.24 & 33.91; 5.12-5.53 \\ 
					0 & 0 & 0.8 & 0.4 & 7.99; 3.36-3.45 & 13.61; 3.67-3.86 & 19.09; 3.90-4.23 & 24.26; 4.06-4.51 & 13.38; 4.42-4.60 & 21.47; 4.78-5.07 & 27.68; 5.03-5.46 & 33.91; 5.23-5.74 \\ 
					1 & 0 & 0.2 & 0 & 49.93; 4.37-4.49 & 99.82; 4.78-4.98 & 149.66; 5.03-5.31 & 199.50; 5.21-5.64 & 50.00; 5.01-5.06 & 100.00; 5.41-5.49 & 150; 5.65-5.74 & 200.00; 5.82-5.92 \\ 
					1 & 0 & 0.2 & 0.4 & 49.93; 4.39-4.58 & 99.82; 4.83-5.25 & 149.66; 5.12-6.02 & 199.50; 5.46-7.06 & 50.00; 5.03-5.09 & 100.00; 5.43-5.52 & 150; 5.67-5.78 & 200.00; 5.84-5.99 \\ 
					1 & 0 & 0.5 & 0 & 37.22; 4.36-5.00 & 69.50; 4.96-6.66 & 100.75; 5.49-8.08 & 130.81; 6.10-9.57 & 47.90; 5.16-5.48 & 93.54; 5.75-6.53 & 137.87; 6.13-7.55 & 181.67; 6.5-8.69 \\ 
					1 & 0 & 0.5 & 0.4 & 37.22; 4.46-5.61 & 69.50; 5.22-7.91 & 100.75; 6.07-10.05 & 130.81; 7.09-12.42 & 47.90; 5.26-5.89 & 93.54; 5.98-7.82 & 137.87; 6.74-9.8 & 181.67; 7.63-11.83 \\ 
					1 & 0 & 0.8 & 0 & 7.99; 3.35-3.51 & 13.61; 3.66-4.01 & 19.09; 3.87-4.50 & 24.26; 4.02-4.92 & 13.38; 4.38-4.64 & 21.47; 4.70-5.30 & 27.68; 4.92-5.76 & 33.91; 5.12-6.28 \\ 
					1 & 0 & 0.8 & 0.4 & 7.99; 3.36-3.59 & 13.61; 3.67-4.10 & 19.09; 3.90-4.65 & 24.26; 4.06-5.13 & 13.38; 4.42-6.00 & 21.47; 4.78-5.61 & 27.68; 5.03-6.17 & 33.91; 5.23-6.72 \\ 
					5 & 0 & 0.2 & 0 & 49.93; 4.37-10.26 & 99.82; 4.78-18.56 & 149.66; 5.03-26.91 & 199.50; 5.21-35.30 & 50.00; 5.01-6.85 & 100.00; 5.41-9.83 & 150; 5.65-13.89 & 200.00; 5.82-18.07 \\ 
					5 & 0 & 0.2 & 0.4 & 49.93; 4.39-11.20 & 99.82; 4.83-20.72 & 149.66; 5.12-29.94 & 199.50; 5.46-39.25 & 50.00; 5.03-7.04 & 100.00; 5.43-10.66 & 150; 5.67-15.22 & 200.00; 5.84-19.75 \\ 
					5 & 0 & 0.5 & 0 & 37.22; 4.36-19.07 & 69.50; 4.96-31.31 & 100.75; 5.49-42.12 & 130.81; 6.10-53.13 & 47.90; 5.16-18.04 & 93.54; 5.75-29.97 & 137.87; 6.13-41.02 & 181.67; 6.5-52.18 \\ 
					5 & 0 & 0.5 & 0.4 & 37.22; 4.46-21.71 & 69.50; 5.22-36.12 & 100.75; 6.07-48.70 & 130.81; 7.07-61.79 & 47.90; 5.26-20.68 & 93.54; 5.98-34.64 & 137.87; 6.74-47.27 & 181.67; 7.63-60.01 \\ 
					5 & 0 & 0.8 & 0 & 7.99; 3.35-7.62 & 13.61; 3.66-11.06 & 19.09; 3.87-14.46 & 24.26; 4.02-17.14 & 13.38; 4.38-12.82 & 21.47; 4.70-17.65 & 27.68; 4.92-20.76 & 33.91; 5.12-23.91 \\ 
					5 & 0 & 0.8 & 0.4 & 7.99; 3.36-7.89 & 13.61; 3.67-11.5 & 19.09; 3.90-15.04 & 24.26; 4.06-18.06 & 13.38; 4.42-13.73 & 21.47; 4.78-19.05 & 27.68; 5.03-22.34 & 33.91; 5.23-25.85 \\ 
					0 & 0.15 & 0.2 & 0 & 49.43; 4.42-4.48 & 98.56; 4.87-4.98 & 147.47; 5.17-5.35 & 196.35; 5.42-5.70 & 49.99; 5.04-5.05 & 100.00; 5.46-5.48 & 149.99; 5.72-5.74 & 199.99; 5.90-5.93 \\ 
					0 & 0.15 & 0.2 & 0.4 & 49.43; 4.46-4.55 & 98.56; 4.99-5.21 & 147.47; 5.48-5.95 & 196.35; 6.15-6.93 & 49.99; 5.07-5.08 & 100.00; 5.49-5.51 & 149.99; 5.75-5.79 & 199.99; 5.95-6.00 \\ 
					0 & 0.15 & 0.5 & 0 & 29.55; 4.23-4.43 & 54.59; 4.86-5.32 & 78.28; 5.55-6.30 & 101.42; 6.33-7.27 & 43.06; 5.16-5.25 & 81.86; 5.75-5.98 & 118.39; 6.23-6.64 & 154.1; 6.75-7.30 \\ 
					0 & 0.15 & 0.5 & 0.4 & 29.55; 4.32-4.62 & 54.59; 5.09-5.81 & 78.28; 5.90-7.05 & 101.42; 6.94-8.48 & 43.06; 5.30-5.48 & 81.86; 6.12-6.62 & 118.39; 7.04-7.87 & 154.1; 7.98-9.10 \\ 
					0 & 0.15 & 0.8 & 0 & 2.29; 1.97-1.97 & 3.70; 2.72-2.74 & 4.86; 3.01-3.03 & 6.39; 3.22-3.25 & 5.01; 3.75-3.76 & 8.76; 4.13-4.16 & 12.25; 4.34-4.39 & 15.79; 4.53-4.62 \\ 
					0 & 0.15 & 0.8 & 0.4 & 2.29; 1.97-1.98 & 3.70; 2.73-2.75 & 4.86; 3.02-3.04 & 6.39; 3.22-3.26 & 5.01; 3.75-3.77 & 8.76; 4.14-4.20 & 12.25; 4.38-4.46 & 15.79; 4.58-4.70 \\ 
					0 & 0.3 & 0.2 & 0 & 47.7; 4.44-4.52 & 94.62; 4.96-5.12 & 141.32; 5.37-5.65 & 187.8; 5.79-6.21 & 49.94; 5.06-5.07 & 99.89; 5.51-5.53 & 149.82; 5.78-5.81 & 199.77; 5.97-6.01 \\
					0 & 0.3 & 0.2 & 0.4 & 47.7; 4.51-4.63 & 94.62; 5.22-5.53 & 141.32; 5.92-6.48 & 187.8; 6.85-7.66 & 49.94; 5.10-5.11 & 99.89; 5.55-5.58 & 149.82; 5.84-5.89 & 199.77; 6.10-6.17 \\ 
					0 & 0.3 & 0.5 & 0 & 20.72; 4.00-4.12 & 38.71; 4.62-4.89 & 56.55; 5.28-5.76 & 74.18; 6.12-6.69 & 35.58; 5.07-5.14 & 66.92; 5.69-5.84 & 96.03; 6.23-6.51 & 124.39; 6.84-7.21 \\ 
					0 & 0.3 & 0.5 & 0.4 & 20.72; 4.08-4.23 & 38.71; 4.80-5.17 & 56.55; 5.49-6.10 & 74.18; 6.43-7.30 & 35.58; 5.23-5.34 & 66.92; 6.09-6.39 & 96.03; 7.04-7.55 & 124.39; 7.95-8.61 \\ 
					\hline
			  \end{tabular}
		\end{adjustbox}\caption{Each entrance of the Table contains the mean dimension $p^*$ of the variance and covariance matrix estimate and the mean values of $\lambda$ and $\lambda_0$ for different values of $n$, $p$, $c_1$, $c_2$, $\pi_{j1}$, $\rho$ for $K = 2$. The presented results are the average of $1000$ repetitions.}
			\label{Tab::DimLambdaSIM-K2}
		\end{minipage}
	}
	
\end{table}

\begin{table}[!h]
\rotatebox{270}{
	\begin{minipage}{1.45\textwidth}
		\centering
		\begin{adjustbox}{max width=\textwidth} 
				\begin{tabular}{cccc|cccc|cccc}
					& & & &\multicolumn{4}{c|}{$n_k = 5$} & \multicolumn{4}{c}{$n_k = 10$}\\
					$c_1$ & $c_2$ & $\pi_{j1}$ & $\rho$ & \multicolumn{1}{c}{$p = 50$} & \multicolumn{1}{c}{$p = 100$} & \multicolumn{1}{c}{$p = 150$} & \multicolumn{1}{c|}{$p = 200$} & \multicolumn{1}{c}{$p = 50$} & \multicolumn{1}{c}{$p = 100$} & \multicolumn{1}{c}{$p = 150$} & \multicolumn{1}{c}{$p = 200$}\\ 
					\hline
					0 & 0 & 0.2 & 0 & 49.94; 5.00-5.01 & 99.87; 5.39-5.42 & 149.80; 5.62-5.66 & 199.74; 5.79-5.83 & 50.00; 5.66-5.67 & 100.00; 6.04-6.04 & 150.00; 6.27-6.27 & 200.00; 6.43-6.44 \\ 
					0 & 0 & 0.2 & 0.4 & 49.94; 5.01-5.04 & 99.87; 5.40-5.44 & 149.80; 5.63-5.68 & 199.74; 5.80-5.86 & 50.00; 5.68-5.69 & 100; 6.06-6.07 & 150; 6.28-6.30 & 200.00; 6.45-6.46 \\ 
					0 & 0 & 0.5 & 0 & 42.79; 4.98-5.12 & 84.38; 5.44-5.72 & 124.98; 5.74-6.17 & 165.29; 5.96-6.53 & 49.76; 5.72-5.75 & 99.57; 6.18-6.25 & 149.33; 6.49-6.59 & 199.05; 6.72-6.84 \\ 
					0 & 0 & 0.5 & 0.4 & 42.79; 5.03-5.24 & 84.38; 5.50-6.03 & 124.98; 5.84-6.91 & 165.29; 6.15-7.83 & 49.76; 5.78-5.83 & 99.57; 6.25-6.35 & 149.33; 6.55-6.70 & 199.05; 6.80-7.00 \\ 
					0 & 0 & 0.8 & 0 & 5.38; 3.74-3.78 & 9.33; 4.10-4.18 & 12.65; 4.27-4.41 & 15.87; 4.39-4.57 & 16.29; 5.12-5.23 & 26.61; 5.41-5.58 & 35.56; 5.59-5.85 & 43.18; 5.71-6.07 \\
					0 & 0 & 0.8 & 0.4 & 5.38; 3.75-3.82 & 9.33; 4.11-4.26 & 12.65; 4.28-4.49 & 15.87; 4.40-4.65 & 16.29; 5.15-5.35 & 26.61; 5.46-5.86 & 35.56; 5.66-6.25 & 43.18; 5.81-6.61 \\ 
					1 & 0 & 0.2 & 0 & 49.94; 5.00-5.04 & 99.87; 5.39-5.46 & 149.8; 5.62-5.70 & 199.74; 5.79-5.88 & 50.00; 5.66-5.68 & 100.00; 6.04-6.07 & 150.00; 6.27-6.30 & 200.00; 6.43-6.48 \\ 
					1 & 0 & 0.2 & 0.4 & 49.94; 5.01-5.07 & 99.87; 5.40-5.48 & 149.8; 5.63-5.74 & 199.74; 5.80-5.93 & 50.00; 5.68-5.71 & 100.00; 6.06-6.09 & 150.00; 6.28-6.33 & 200.00; 6.45-6.5. \\ 
					1 & 0 & 0.5 & 0 & 42.79; 4.98-5.22 & 84.38; 5.44-5.94 & 124.98; 5.74-6.56 & 165.29; 5.96-7.18 & 49.76; 5.72-5.81 & 99.57; 6.18-6.35 & 149.33; 6.49-6.71 & 199.05; 6.72-6.99 \\ 
					1 & 0 & 0.5 & 0.4 & 42.79; 5.03-5.45 & 84.38; 5.50-6.60 & 124.98; 5.84-7.95 & 165.29; 6.15-9.46 & 49.76; 5.78-5.91 & 99.57; 6.25-6.49 & 149.33; 6.55-6.90 & 199.05; 6.80-7.39 \\ 
					1 & 0 & 0.8 & 0 & 5.38; 3.74-3.80 & 9.33; 4.10-4.24 & 12.65; 4.27-4.49 & 15.87; 4.39-4.69 & 16.29; 5.12-5.31 & 26.61; 5.41-5.78 & 35.56; 5.59-6.14 & 43.18; 5.71-6.48 \\ 
					1 & 0 & 0.8 & 0.4 & 5.38; 3.75-3.87 & 9.33; 4.11-4.37 & 12.65; 4.28-4.62 & 15.87; 4.40-4.85 & 16.29; 5.15-5.52 & 26.61; 5.46-6.28 & 35.56; 5.66-6.86 & 43.18; 5.81-7.36 \\ 
					5 & 0 & 0.2 & 0 & 49.94; 5-6.24 & 99.87; 5.39-8.20 & 149.80; 5.62-10.85 & 199.74; 5.79-13.82 & 50.00; 5.66-6.24 & 100.00; 6.04-6.99 & 150.00; 6.27-7.61 & 200.00; 6.43-8.27 \\ 
					5 & 0 & 0.2 & 0.4 & 49.94; 5.01-6.35 & 99.87; 5.40-8.66 & 149.80; 5.63-11.61 & 199.74; 5.80-14.88 & 50.00; 5.68-6.24 & 100.00; 6.06-7.01 & 150.00; 6.28-7.72 & 200.00; 6.45-8.55 \\ 
					5 & 0 & 0.5 & 0 & 42.79; 4.98-14.81 & 84.38; 5.44-24.16 & 124.98; 5.74-32.34 & 165.29; 5.96-40.26 & 49.76; 5.72-10.28 & 99.57; 6.18-17.01 & 149.33; 6.49-22.83 & 199.05; 6.72-28.37 \\ 
					5 & 0 & 0.5 & 0.4 & 42.79; 5.03-15.92 & 84.38; 5.5-26.36 & 124.98; 5.84-35.23 & 165.29; 6.15-44.29 & 49.76; 5.78-11.08 & 99.57; 6.25-18.4 & 149.33; 6.55-24.87 & 199.05; 6.8-31.28 \\ 
					5 & 0 & 0.8 & 0 & 5.38; 3.74-5.98 & 9.33; 4.10-8.61 & 12.65; 4.27-10.72 & 15.87; 4.39-12.34 & 16.29; 5.12-13.82 & 26.61; 5.41-18.91 & 35.56; 5.59-22.55 & 43.18; 5.71-24.99 \\ 
					5 & 0 & 0.8 & 0.4 & 5.38; 3.75-6.25 & 9.33; 4.11-9.18 & 12.65; 4.28-11.20 & 15.87; 4.40-12.98 & 16.29; 5.15-14.92 & 26.61; 5.46-20.54 & 35.56; 5.66-24.33 & 43.18; 5.81-27.09 \\ 
					0 & 0.15 & 0.2 & 0 & 49.54; 5.01-5.04 & 99.1; 5.42-5.48 & 148.68; 5.67-5.74 & 198.23; 5.85-5.93 & 50.00; 5.67-5.68 & 100.00; 6.06-6.07 & 150.00; 6.3-6.32 & 199.99; 6.48-6.50 \\ 
					0 & 0.15 & 0.2 & 0.4 & 49.54; 5.04-5.08 & 99.10; 5.45-5.51 & 148.68; 5.69-5.79 & 198.23; 5.88-5.99 & 50.00; 5.70-5.71 & 100.00; 6.09-6.11 & 150.00; 6.33-6.35 & 199.99; 6.50-6.53 \\ 
					0 & 0.15 & 0.5 & 0 & 34.27; 4.85-5.01 & 66.14; 5.29-5.61 & 96.47; 5.58-6.10 & 125.47; 5.79-6.55 & 48.61; 5.73-5.78 & 96.66; 6.22-6.34 & 144.60; 6.54-6.72 & 192.25; 6.81-7.04 \\ 
					0 & 0.15 & 0.5 & 0.4 & 34.27; 4.89-5.14 & 66.14; 5.36-6.00 & 96.47; 5.69-6.82 & 125.47; 5.97-7.75 & 48.61; 5.79-5.89 & 96.66; 6.30-6.50 & 144.60; 6.65-7.03 & 192.25; 7.00-7.71 \\ 
					0 & 0.15 & 0.8 & 0 & 1.38; 1.05-1.05 & 1.63; 1.53-1.54 & 1.90; 1.98-1.98 & 2.23; 2.40-2.41 & 3.30; 3.79-3.81 & 5.33; 4.39-4.41 & 6.87; 4.62-4.65 & 8.58; 4.77-4.83 \\ 
					0 & 0.15 & 0.8 & 0.4 & 1.38; 1.05-1.05 & 1.63; 1.53-1.54 & 1.90; 1.98-1.98 & 2.23; 2.4-2.41 & 3.30; 3.79-3.83 & 5.33; 4.40-4.45 & 6.87; 4.62-4.68 & 8.58; 4.79-4.89 \\
					0 & 0.3 & 0.2 & 0 & 47.74; 5.01-5.05 & 95.66; 5.44-5.52 & 143.3; 5.70-5.80 & 191.15; 5.88-6.02 & 49.93; 5.68-5.69 & 99.88; 6.08-6.10 & 149.84; 6.34-6.36 & 199.78; 6.52-6.56 \\ 
					0 & 0.3 & 0.2 & 0.4 & 47.74; 5.04-5.10 & 95.66; 5.47-5.57 & 143.3; 5.73-5.89 & 191.15; 5.94-6.16 & 49.93; 5.72-5.73 & 99.88; 6.12-6.15 & 149.84; 6.37-6.41 & 199.78; 6.56-6.60 \\ 
					0 & 0.3 & 0.5 & 0 & 20.73; 4.57-4.67 & 38.78; 4.95-5.17 & 55.54; 5.20-5.52 & 71.67; 5.40-5.85 & 41.64; 5.66-5.73 & 81.10; 6.14-6.30 & 120.36; 6.48-6.74 & 158.43; 6.72-7.09 \\ 
					0 & 0.3 & 0.5 & 0.4 & 20.73; 4.60-4.77 & 38.78; 5.01-5.39 & 55.54; 5.28-5.85 & 71.67; 5.53-6.44 & 41.64; 5.72-5.85 & 81.10; 6.24-6.57 & 120.36; 6.67-7.33 & 158.43; 7.11-8.22 \\ 
					\hline
			 \end{tabular}
		\end{adjustbox}\caption{Each entrance of the Table contains the mean dimension $p^*$ of the variance and covariance matrix estimate and the mean values of $\lambda$ and $\lambda_0$ for different values of $n$, $p$, $c_1$, $c_2$, $\pi_{j1}$, $\rho$ for $K = 4$. The presented results are the average of $1000$ repetitions.}
			\label{Tab::DimLambdaSIM-K4}
		\end{minipage}
	}
	
\end{table}

\end{document}